\documentclass[sigconf]{acmart}

\AtBeginDocument{%
  }

\copyrightyear{2025}
\acmYear{2025}
\setcopyright{acmlicensed}\acmConference[KDD '25]{Proceedings of the 31st ACM SIGKDD Conference on Knowledge Discovery and Data Mining V.2}{August 3--7, 2025}{Toronto, ON, Canada}
\acmBooktitle{Proceedings of the 31st ACM SIGKDD Conference on Knowledge Discovery and Data Mining V.2 (KDD '25), August 3--7, 2025, Toronto, ON, Canada}
\acmDOI{10.1145/3711896.3737233}
\acmISBN{979-8-4007-1454-2/2025/08}

\usepackage[utf8]{inputenc}
\usepackage[T1]{fontenc}    
\usepackage{hyperref}        
\usepackage{url}            
\usepackage{booktabs}       
\usepackage{amsfonts}      
\usepackage{nicefrac}       
\usepackage{microtype}     
\usepackage{xcolor}         
\usepackage{graphicx}
\usepackage{amsmath}
\usepackage{multirow}
\usepackage{algorithm}
\usepackage{algorithmic}

\begin{document}

\title{Hierarchical Lexical Graph for Enhanced Multi-Hop Retrieval}

\author{Abdellah Ghassel}
\orcid{0009-0007-3042-9747}
\authornote{Equal contribution.}
\authornote{Work done during internship at Amazon.}
\affiliation{
  \institution{Queen's University}
  \city{Kingston}
  \country{Canada}
}
\email{abdellah.ghassel@queensu.ca}

\author{Ian Robinson}
\orcid{0009-0000-4247-8588}
\authornotemark[1]
\affiliation{
  \institution{Amazon}
  \city{London}
  \country{England}
}
\email{ianrob@amazon.co.uk}

\author{Gabriel Tanase}
\orcid{0009-0002-6906-6983}
\affiliation{
  \institution{Amazon}
  \city{Seattle}
  \country{USA}
}
\email{igtanase@amazon.com}

\author{Hal Cooper}
\orcid{0009-0007-3168-175X}
\affiliation{
  \institution{Amazon}
  \city{Seattle}
  \country{USA}
}
\email{halcoope@amazon.com}

\author{Bryan Thompson}
\orcid{0009-0008-5782-236X}
\affiliation{
  \institution{Amazon}
  \city{Seattle}
  \country{USA}
}
\email{bryant@amazon.com}

\author{Zhen Han}
\orcid{0009-0003-8845-0507}
\affiliation{
  \institution{Amazon}
  \city{Santa Clara}
  \country{USA}
}
\email{zhenhz@amazon.com}

\author{Vassilis N. Ioannidis}
\orcid{0000-0002-8367-0733}
\affiliation{
  \institution{Amazon}
  \city{Santa Clara}
  \country{USA}
}
\email{ivasilei@amazon.com}

\author{Soji Adeshina}
\orcid{0000-0003-3945-3640}
\affiliation{
  \institution{Amazon}
  \city{Santa Clara}
  \country{USA}
}
\email{adesojia@amazon.com}

\author{Huzefa Rangwala}
\orcid{0000-0003-0435-0035}
\affiliation{
  \institution{Amazon}
  \city{Santa Clara}
  \country{USA}
}
\email{rhuzefa@amazon.com}

\renewcommand{\shortauthors}{Abdellah Ghassel et al.}

\begin{abstract}
Retrieval-Augmented Generation (RAG) grounds large language models in external evidence, yet it still falters when answers must be pieced together across semantically distant documents. We close this gap with the Hierarchical Lexical Graph (HLG), a three-tier index that (i) traces every atomic proposition to its source, (ii) clusters propositions into latent topics, and (iii) links entities and relations to expose cross-document paths. On top of HLG we build two complementary, plug-and-play retrievers: StatementGraphRAG, which performs fine-grained entity-aware beam search over propositions for high-precision factoid questions, and TopicGraphRAG, which selects coarse topics before expanding along entity links to supply broad yet relevant context for exploratory queries.
Additionally, existing benchmarks lack the complexity required to rigorously evaluate multi-hop summarization systems, often focusing on single-document queries or limited datasets. To address this, we introduce a synthetic dataset generation pipeline that curates realistic, multi-document question-answer pairs, enabling robust evaluation of multi-hop retrieval systems. Extensive experiments across five datasets demonstrate that our methods outperform naive chunk-based RAG, achieving an average relative improvement of 23.1\% in retrieval recall and correctness. \footnote{Open-source Python library is available at \texttt{github.com/awslabs/graphrag-toolkit}}
\end{abstract}


  
\maketitle

\section{Introduction}

Retrieval-Augmented Generation (RAG) systems have gained attention for enhancing large language models (LLMs) with external knowledge, enabling more grounded and accurate responses to complex questions \citep{lewis2021rag, Gao2023RetrievalAugmentedGF}. Despite these advances, current RAG systems face significant challenges with multi-hop reasoning, where answers may be synthesized from multiple, semantically diverse text segments or documents \citep{Tang2024MultiHopRAGBR}. For example, consider the query: \textit{‘How did the FTC lawsuit affect the stock of one of the leading e-commerce companies?’} This requires retrieving facts about the lawsuit, Amazon's financial performance, and market reactions, likely dispersed across multiple, unconnected documents.

Most RAG systems struggle in such scenarios because they rely primarily on vector similarity search (VSS). While VSS excels at finding texts that closely match the query's surface-level semantics, it often fails to bridge related but contextually distant pieces of information \citep{besta2024multihead}. In the previous example, connecting ‘company lawsuit details’ from one document with ‘stock performance data’ in another may require structured links or graph-based expansions, as they lack obvious semantic ties. Graph-based retrieval strategies provide a promising solution as they model explicit edges, such as shared entities or discourse links, thereby overcoming the narrowness of vector-based methods \citep{Fang2019HierarchicalGN, graphAttention, gcnn}.

In addition, we emphasize the importance of retrieval-unit granularity. Conventional systems often utilize large text chunks, consisting of various sentences, leading to the retrieval of extraneous information. To address this, \citet{chen2024dense} advocate for smaller retrieval units, such as atomic propositions, to enhance precision in information retrieval.
 
Building on these insights, we propose a Hierarchical Lexical Graph (HLG) framework to address the gap between surface-level similarity and structured multi-hop evidence. While prior work emphasizes the benefits of finer retrieval units \citep{chen2024dense}, HLG extends these ideas by integrating three interconnected tiers: Lineage, Summarization, and Entity-Relationship. Specifically, HLG (1) preserves the lineage of each statement for accurate provenance, (2) clusters statements around topics for flexible retrieval, and (3) links entities and relationships to enable bottom-up graph traversal. This multi-tier design enables more precise retrieval of diverse facts, even when they share minimal semantic overlap.

Using this framework, we propose two complementary RAG methods tailored to different retrieval needs:
\begin{itemize}
\item \textbf{StatementGraphRAG} focuses on individual propositions in the Summarization Tier. It links them across documents using the Entity-Relationship Tier and preserves provenance via the Lineage Tier. This is ideal for detailed queries needing high-precision evidence (e.g., factoid questions).
\item \textbf{TopicGraphRAG} retrieves clusters of statements (i.e., topical groups) from the Summarization Tier, using entity relationships to connect thematically related clusters and relying on the Lineage Tier for source traceability. This is efficient for broader, open-ended or higher-level queries.
\end{itemize}

The choice depends on query type: StatementGraphRAG for specific, single-answer queries, and TopicGraphRAG for exploratory or multi-faceted questions.

Despite the promise of these approaches, a major obstacle remains: most existing QA datasets fail to demand true multi-hop retrieval, either restricting questions to single documents or offering a small set of queries. WikiHowQA \citep{2023-wikihowqa}, for example, focuses on single-document responses, while the SEC10Q \citep{docugami_kg_rag_2023} dataset contains only 195 queries. To address this gap, we present a synthetic multi-hop summarization pipeline using HLG as a backbone. We generate 674 question-answer pairs from the MultiHop-RAG corpus \citep{Tang2024MultiHopRAGBR}, then apply a second-pass LLM critique and automated retrieval checks \citep{ru2024ragchecker} to ensure each question genuinely requires multi-hop inference and has sufficient supporting evidence.

Our paper makes the following contributions:
\begin{enumerate}
    \item We introduce StatementGraphRAG and TopicGraphRAG, advancing multi-hop QA with fine-grained and structured retrieval capabilities.
    \item We develop a pipeline synthesizing high-quality multi-hop summarization queries, closing gaps in existing benchmarks.
    \item Through extensive experiments across diverse datasets, we show significant performance gains over chunk-based RAG baselines.
\end{enumerate}

\section{Related Work}
\label{Related Work}

\subsection{Multi-hop Reasoning with Graphs}
Graph-based approaches have been actively used in modelling relationships required for effective multi-hop reasoning. \citet{Fang2019HierarchicalGN} introduced the Hierarchical Graph Network (HGN) for multi-hop question answering, which integrates nodes of varying granularity including questions, paragraphs, sentences, and entities using graph neural networks. This facilitates joint prediction of answers and supporting facts by enabling multi-hop information propagation across different levels of the graph. However, HGN faces significant limitations that hinder its scalability and generalizability. First, the storage of embeddings across multiple granularities incurs substantial memory and computational overhead, particularly for large datasets. Second, its reliance on explicit relational structures, such as Wikipedia hyperlinks, restricts its applicability to domains where such curated connections are sparse. Our HLG extends HGN's foundational concepts by embedding both fine-grained propositions and higher-level topics within a unified graph structure.

\citet{edge2024localglobalgraphrag} proposed GraphRAG, a graph-indexing method designed for global summarization tasks. By constructing an entity-relation knowledge graph and hierarchically detecting communities on the graph, GraphRAG enables summarization at indexing time. However, its reliance on pre-defined entities and static Leiden-generated communities \citep{Traag_2019} poses limitations for dynamically expanding domains. Incorporating additional data requires recalculating the community structure for the entire graph, which can be computationally expensive. In contrast, HLG is designed for continuous data ingestion and supports multiple granularity tiers, facilitating more efficient updates and robust retrieval. Rather than relying on pre-computed communities, HLG dynamically retrieves relevant topics at query time, enabling greater scalability and resource efficiency.

\subsection{Dense Retrieval Granularity}
\citet{chen2024dense} analyze how the granularity of retrieval units affects dense retrieval. Their findings indicate that fine-grained propositions outperform larger units (e.g., sentences or passages) in both retrieval precision and downstream task performance. While \citet{chen2024dense} focus on dense retrieval for standard datasets, we extend this principle by integrating hierarchical graph structures to ensure relevance and connectivity across diverse sources. To enhance the adaptability of statement extraction for structured and tabular data (such as SEC filings), an LLM-modified version of the propositionizer (\texttt{propositionizer-wiki-flan-t5-large}) was incorporated.

\section{Hierarchical Lexical Graph}
The design of our RAG solution is driven by a systematic "working backward" approach that, given a workload's question-answering needs, identifies optimal retrieval and generation strategies and appropriate indexing and storage systems in support of those needs. In this design, consideration is given to the types of end-user or application-specific data needs the workflow is intended to support, the data required to meet these needs, and the indexing structures best suited for efficient retrieval. HLG is constructed once per dataset during initial indexing to optimize relevance for the target domain. However, HLG supports incremental ingestion of new documents and entity propagation without full reindexing, due to modular updates in its tiers (Section~\ref{sec:hlg_tiers}).

\begin{figure}
    \centering
    \includegraphics[width=1\linewidth]{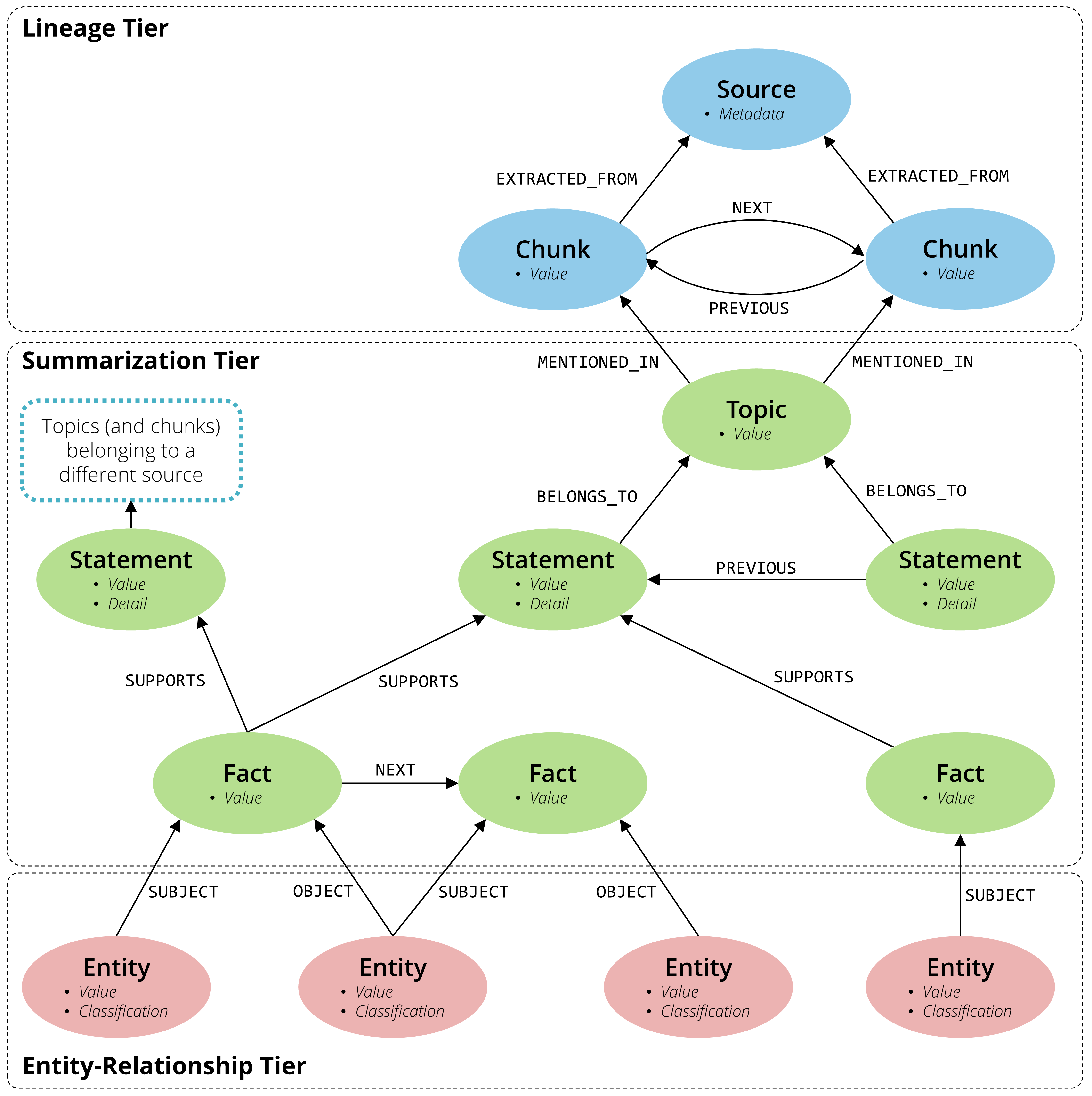}
    \caption{Hierarchical Lexical Graph Model. Statements are atomic propositions \textit{(e.g., "Company X acquired Company Y")}. Topics, thematic groups \textit{(e.g., "Q3 Financial Results")}. Facts, structured S-P-O triplets \textit{(e.g., <Company X, acquired, Company Y>)}. Entities, named concepts \textit{(e.g., "Company X")}. Keywords (not shown but used in retrieval) are salient terms extracted from queries \textit{(e.g., "acquisition", "Company X")}.}
    \Description{A coloured model showing the components of the three tiers of the lexical graph model}
    \label{fig:lexicalGraphModel}
\end{figure}

\subsection{Statement Retrieval and Graph Structures}
\label{sec:hlg_tiers}
Determining the optimal size and structure of retrieval units is critical in search-based workflows. Traditional RAG systems often rely on text "chunks" (segments larger than sentences, smaller than documents). However, such coarse-grained units frequently amalgamate unrelated sentences, reducing precision. In contrast, our system centers on "propositions" (standalone assertions extracted from text) as the primary retrieval unit. This granularity allows for precise, contextually relevant inputs for LLMs.

The retrieval system is built upon a lexical graph model, illustrated in Figure~\ref{fig:lexicalGraphModel}, comprising three interconnected tiers that enable fine-grained connectivity across documents.

\subsubsection{Lineage Tier} Establishes the foundation of the graph, ensuring traceability and contextual integrity.
\begin{itemize}
    \item \textbf{Source Nodes}: Metadata such as document origin, date, and author information, for provenance.
    \item \textbf{Chunk Nodes}: Sequentially linked text segments, preserving context for further analysis and maintaining lineage.
\end{itemize}

This tier is useful in compliance-driven scenarios, where maintaining the original context and source of retrieved information is critical for interpretability.

\subsubsection{Entity-Relationship Tier} Captures relationships between entities, serves as an entry point for structured, keyword-based searches.
\begin{itemize}
    \item \textbf{Entity Nodes}: Key entities (e.g., "Amazon") classified by category (e.g., "Company").
    \item \textbf{Relationship Edges}: Inter-entity relationships (e.g., "FILED LAWSUIT AGAINST"), for structured retrieval tasks.
\end{itemize}

This tier supports complex aggregation queries, such as \textit{"Which companies filed lawsuits in Q3 2023?"}, while dynamically incorporating new classifications to enhance adaptability.

\subsubsection{Summarization Tier} Links granular facts and statements to broader topics, forming hierarchical semantic units.
\begin{itemize}
    \item \textbf{Facts}: Discrete semantic units (subject-predicate-object triplets) that connect granular and global insights.
    \item \textbf{Statements}: Propositions extracted from source documents, forming the backbone of the retrieval process.
    \item \textbf{Topics}: Thematic summaries grouping related statements, enhancing intra-document connectivity.
\end{itemize}

By connecting statements to overarching topics, this tier facilitates both local and global reasoning, ensuring efficient retrieval strategies for multi-hop QA tasks.

\subsection{Graph Connectivity}
HLG optimizes retrieval through a hybrid approach that combines semantic similarity and graph traversal, ensuring flexible and efficient navigation across the graph:
\begin{enumerate}
    \item \textbf{Local/Global Connectivity}: Topics provide intra-document connectivity by linking statements within a source. In contrast, facts enable inter-document connections, ensuring comprehensive retrieval for complex queries.
    \item \textbf{Vector-Based Entry Points}: Both topics and/or statements are embedded for vector-based searches, enabling precise semantic similarity matching. Topic embeddings incorporate associated statements to enhance query alignment.
    \item \textbf{Keyword-Based Entry Points}: Query keywords can be matched with entities in the Entity-Relationship Tier, essential for bottom-up lookups.
\end{enumerate}

\section{Proposed Retrieval Models}
Two methods leverage HLG: StatementGraphRAG (Section~\ref{subsec:statementgraphrag}) and TopicGraphRAG (Section~\ref{subsec:topicgraphrag}).
The choice between them depends on the query's nature. StatementGraphRAG excels at precise, fact-based queries seeking specific information, often with a single expected answer. Its fine-grained statement retrieval and entity linking ensure high precision. TopicGraphRAG is better for broader, exploratory queries or those requiring synthesis across multiple themes. Its topic-level retrieval provides wider context, suitable for open-ended questions or summarization tasks.

\subsection{StatementGraphRAG}
\label{subsec:statementgraphrag}

\noindent
\textbf{Overview.} 
The StatementGraphRAG pipeline has four components, designed to progressively refine search results:
\begin{enumerate}
    \item \emph{Keyword-Based Retrieval}: Extracts query terms and retrieves statements with those terms via explicit entity matching.
    \item \emph{Vector Similarity Search}: Retrieves statements semantically similar to the query.
    \item \emph{Graph Beam Search}: Explores multi-hop neighbours in HLG by traversing shared entities, and scoring resulting paths.
    \item \emph{Reranking}: Rescores all candidates to finalize top-$k$ results.
\end{enumerate}

\vspace{0.5em}
\noindent
\textbf{Notation.}
Let $Q\in\mathcal{D}$ be the user query and
\[
\mathbf{e}_Q \;:=\; \operatorname{Embed}(Q)\in\mathbb{R}^{d}
\]
its $d$-dimensional embedding.  
Let $\mathcal{G}=(V,E)$ be the lexical graph and
$\mathcal{S}_{\!\mathcal{G}}\subseteq V$ the set of statement nodes.
Define the set of extracted keywords (and synonyms):
\begin{equation}
  K \;=\; \{\,k_i \mid k_i \text{ extracted from } Q\}, \qquad i=1,\dots,N_k.
  \label{eq:keywords}
\end{equation}

\subsubsection*{Step 1: Keyword Retrieval}
For $s\in\mathcal{S}_{\!\mathcal{G}}$, let
\(
  K_s \subseteq K
\)
be the keywords whose entities appear in~$s$.
We rank statements by:
\begin{equation}
  \operatorname{Score}_{\text{kw}}(s) \;=\; |K_s|,
  \label{eq:score_keywords}
\end{equation}
breaking ties with cosine similarity
\(
  \operatorname{sim}(\mathbf{e}_Q,\mathbf{e}_s)
\).
The $k$ highest-scoring statements form:
\begin{equation}
  \mathcal{S}_{\text{kw}}
  \;=\;
  \operatorname{Top}_k\!\bigl(\mathcal{S}_{\!\mathcal{G}},\,
  \operatorname{Score}_{\text{kw}}\bigr).
\end{equation}

\subsubsection*{Step 2: Vector Similarity Search}
In parallel to Step 1, we retrieve:
\begin{equation}
  \mathcal{S}_{\text{vss}}
  \;=\;
  \operatorname{Top}_k\!
    \bigl\{\,s\in\mathcal{S}_{\!\mathcal{G}}
           \mid
           \operatorname{sim}(\mathbf{e}_Q,\mathbf{e}_s)\bigr\}.
\end{equation}
The initial candidate set is the union,
\(
  \mathcal{S}_{\text{init}}
  =
  \mathcal{S}_{\text{kw}}\cup \mathcal{S}_{\text{vss}}.
\)

\subsubsection*{Step 3: Graph Beam Search}
For a statement $s$, define its neighbours by shared entities:
\begin{equation}
  \operatorname{Nbr}(s)
  \;=\;
  \bigl\{\,s'\in\mathcal{S}_{\!\mathcal{G}}
          \mid
          \operatorname{Ent}(s)\cap\operatorname{Ent}(s')\neq\varnothing
  \bigr\}.
  \label{eq:neighbours}
\end{equation}

Given a path $P=\langle s_1,\dots,s_n\rangle$, we compute an
attention-weighted path embedding \citep{bahdanau2016neuralmachinetranslationjointly}:
\begin{align}
  \mathbf{e}_{\text{path}}(P)
    &= \sum_{i=1}^{n}\alpha_i\,\mathbf{e}_{s_i},
    &\alpha_i
    &= \frac{\exp\!\bigl(\operatorname{sim}(\mathbf{e}_Q,\mathbf{e}_{s_i})\bigr)}
            {\sum_{j=1}^{n}\exp\!\bigl(\operatorname{sim}(\mathbf{e}_Q,\mathbf{e}_{s_j})\bigr)}.
    \label{eq:path_embedding}
\end{align}
Path relevance is:
\begin{equation}
  \operatorname{Score}_{\text{beam}}(P)
  \;=\;
  \operatorname{sim}\bigl(\mathbf{e}_Q,\mathbf{e}_{\text{path}}(P)\bigr).
  \label{eq:beam_score}
\end{equation}
Starting from each initial state $s\in\mathcal{S}_{\text{init}}$, 
beam search expands the frontier until either  
(i) no child improves the score or (ii) the maximum depth $D_{\text{max}}$ is reached. At every depth, it keeps only the $B$ highest-scoring paths, ranked by Eq.~\eqref{eq:beam_score}.  
This procedure yields:
\begin{equation}
  \mathcal{S}_{\text{beam}}
  \;=\;
  \bigcup_{s\in\mathcal{S}_{\text{init}}}
  \textsc{BeamSearch}(s,\mathcal{G}, B, D_{\text{max}}).
\end{equation}
The candidate pool after graph exploration is:
\(
  \mathcal{S}_{\text{final}}
  =
  \mathcal{S}_{\text{init}}\cup\mathcal{S}_{\text{beam}}.
\)

\subsubsection*{Step 4: Reranking}
Each $s\in\mathcal{S}_{\text{final}}$ is rescored by a cross-encoder reranker:
\begin{equation}
  \operatorname{Score}_{\text{rank}}(s)
  \;=\;
  \operatorname{sim}\bigl(\operatorname{Rerank}(Q),\operatorname{Rerank}(s)\bigr).
\end{equation}
We return the top-$k$ results:
\begin{equation}
  \mathcal{S}_{\text{top}}
  \;=\;
  \operatorname{Top}_k\!
    \bigl(\mathcal{S}_{\text{final}},\operatorname{Score}_{\text{rank}}\bigr).
\end{equation}

\begin{algorithm}[h]
\caption{\textsc{StatementGraphRAG}}
\label{alg:statementgraphrag}
\begin{algorithmic}[1]
\REQUIRE Query $Q$, lexical graph $\mathcal{G}$, embedding function $\operatorname{Embed}(\cdot)$, reranker $\operatorname{Rerank}$, top-$k$
\ENSURE $\mathcal{S}_{\text{top}}$: top-$k$ statements relevant to $Q$
\STATE \textbf{Extract} keywords \& synonyms $\longrightarrow K$ \hfill(Eq.~\ref{eq:keywords})
\STATE \textbf{Keyword Retrieval:} obtain $\mathcal{S}_{\text{kw}}$ \hfill(Eq.~\ref{eq:score_keywords})
\STATE \textbf{Vector Similarity Search:} obtain $\mathcal{S}_{\text{vss}}$
\STATE $\mathcal{S}_{\text{init}} \;\gets\; \mathcal{S}_{\text{kw}} \cup \mathcal{S}_{\text{vss}}$
\STATE \textbf{Graph Beam Search:}
  \FOR{each $s\in\mathcal{S}_{\text{init}}$}
      \STATE Expand $s$ via $\operatorname{Nbr}(s)$ \hfill(Eq.~\ref{eq:neighbours})
      \STATE Compute attention-weighted path embedding \hfill(Eq.~\ref{eq:path_embedding})
      \STATE Update path score \hfill(Eq.~\ref{eq:beam_score})
  \ENDFOR
\STATE $\mathcal{S}_{\text{beam}} \;\gets$ all statements visited during beam search
\STATE $\mathcal{S}_{\text{final}} \;\gets\; \mathcal{S}_{\text{init}} \cup \mathcal{S}_{\text{beam}}$
\STATE \textbf{Rerank:} compute $\operatorname{Score}_{\text{rank}}(s)$ for each $s\in\mathcal{S}_{\text{final}}$
\STATE $\mathcal{S}_{\text{top}} \;\gets\; \operatorname{Top}_k\bigl(\mathcal{S}_{\text{final}},\operatorname{Score}_{\text{rank}}\bigr)$
\RETURN $\mathcal{S}_{\text{top}}$
\end{algorithmic}
\end{algorithm}

\subsection{TopicGraphRAG}
\label{subsec:topicgraphrag}

\noindent
\textbf{Overview.}
\emph{TopicGraphRAG} integrates both top-down (topic-driven) and bottom-up (entity-driven) retrieval to identify and expand thematically relevant information through multi-hop reasoning. Topics typically map to a small number of chunks (1:n, where n is small), often from the same document. This makes the approach more storage-efficient than retrieving isolated statements, which tend to have a 10:1 chunk-to-statement ratio. The pipeline consists of four stages, balancing broad thematic coverage with precise detail:

\begin{enumerate}
    \item \textbf{Top-Down: Topic Discovery.} 
    Embed the query to identify high-level topics aligned with user intent, then retrieve the relevant statements linked to those topics.
    \item \textbf{Bottom-Up: Entity Exploration.}
    Extract query-related keywords and match them with associated entities in the lexical graph. Then, retrieve their associated statements.
    \item \textbf{Graph Beam Search.}
    Topic- and entity-related statements are merged. A beam search explores additional context in multiple hops, guided by a reranker at each step.
    \item \textbf{Final Rerank \& Truncation.}
    Rescore all candidate statements using a reranker, and select the top-$k$ relevant ones.
\end{enumerate}

\noindent
\textbf{Rationale.} 
Using \emph{topics} (top-down) and \emph{entities} (bottom-up), TopicGraphRAG broadens coverage while preserving precision. Multi-hop traversal explores cross-document links, and the final reranking yields a set of statements well-aligned with the user query.

\subsection{Post-Retrieval Processing}
After performing retrieval using StatementGraphRAG or TopicGraphRAG, post-processing can enhance diversity and clarity.

\subsubsection{Statement Diversity: Reducing Redundancy}
\label{subsec:statement_diversity}

In many real-world scenarios, multiple sources may report overlapping information,  which can constrain the context window. To address this, a diversity filtering procedure identifies similar statements and preserves the highest-scoring representative. Empirical evaluations (Table \ref{tab:combined_results}) show a diversity threshold (\(\tau\)) as low as 0.5\% can yield modest gains with negligible overhead.
\begin{enumerate}
    \item \textbf{Preprocessing:} Numeric tokens are standardized, stop words are removed, and text is lemmatized. This ensures consistent representation of numeric and textual data.
    \item \textbf{Similarity Detection:} Vectorize statements via TF-IDF \citep{tfidf_1972}, and compute cosine similarity to detect redundancies.
    \[
    \text{Sim}(S_i, S_j) = \frac{\sum_{k=1}^n \text{TF-IDF}_{S_i,k} \cdot \text{TF-IDF}_{S_j,k}}
    {\sqrt{\sum_{k=1}^n (\text{TF-IDF}_{S_i,k})^2} \cdot \sqrt{\sum_{k=1}^n (\text{TF-IDF}_{S_j,k})^2}},
    \]
    where \( \text{TF-IDF}_{S_i,k} \) represents the \( k \)-th term in statement \( S_i \), and \( n \) is the total number of terms.
    \item \textbf{Filtering:} Define a \emph{diversity threshold} \(\tau\), which corresponds to \(1 - \text{(similarity threshold)}\). Discard the lower-scoring statement and retain the more informative version:
    \[
    \text{Retain } S_i \;\Leftrightarrow\;
    \forall S_j:\; \text{Sim}(S_i, S_j) > (1 - \tau) \Rightarrow \text{Score}(S_i) > \text{Score}(S_j)
    \]

\end{enumerate}

\subsubsection{Statement Enhancement: Tabular Data Processing}
\label{subsec:statement_enhancement_tabular}

For financial reports or documents with tabular data, this step enriches statements by incorporating context from their text chunks. This is important for sources like SEC-10Q filings, where numeric propositions may lack clarity without additional details. Relevant contextual information, such as column headers and row labels, are appended to clarify the meaning of the data, improving the readability and interpretability of numeric propositions. This reduces ambiguity (e.g., clarifying units or time periods), ensures completeness, and makes statements more accurate.

\section{Synthetic Multi-Hop Summarization}
\label{sec:synthetic_dataset}

\noindent
\textbf{Motivation.} 
Existing multi-hop summarization datasets, such as SEC-10Q or WikiHow, lack the complexity or diversity required to evaluate RAG systems. SEC-10Q has a limited set of 195 comprehensive queries, whereas WikiHow is not multi-hop since the answer is derived from one text segment. To address these shortcomings, we introduce a synthetic dataset generation pipeline that assembles multi-hop queries spanning multiple documents, thereby testing robust retrieval and summarization capabilities.

\subsection{Dataset Characteristics}
\begin{itemize}
  \item \textbf{Complexity.} Each question needs information from two to four documents (“hops”).
  \item \textbf{Realism.} Queries mirror real-world inquiries that span multiple themes, needing deeper reasoning.
  \item \textbf{Structure.} Each entry includes a multi-hop question, a ground truth answer (from multiple chunks) and document snippets.
  Ground truth answers are synthesized from multiple document snippets and validated for factual accuracy.
\end{itemize}

\subsection{Pipeline Architecture}
Our dataset creation process uses a four-stage pipeline:
\begin{enumerate}
  \item \textbf{Topic Collection.}
  From a seed topic, we retrieve semantically related topics from different documents.
  \item \textbf{Chunk Selection.}
  Collect chunks from each relevant topic (3-5 distinct articles) for cohesive context.
  \item \textbf{Query Generation.}
  Prompt an LLM with diverse chunks to generate a multi-hop question. These questions require synthesizing information across articles \textit{(e.g., cause-effect, contrasting perspectives, complementary insights).}
  \item \textbf{Critiquing and Refinement.}
  Queries are refined and validated by a second LLM pass to ensure clarity, coherence, and multi-article coverage. Queries that fail validation, for instance due to insufficient evidence, are discarded.
\end{enumerate}

\noindent \textbf{Quality Control and Filtering.}
In Step 4, we utilize a superior LLM (\texttt{Claude-3.5 Sonnet v2} \citep{anthropic2024claude35}) than the one used for query-generation (\texttt{Claude-3 Sonnet} \citep{anthropic_claude3_2024}), to align better with human judgement \citep{Kim2024Prometheus2A}. We incorporate a two-stage critique:
\begin{itemize}
  \item \emph{Query Refinement:} Each initial query-answer pair is examined to enhance clarity, address ambiguities, and enrich the multi-hop nature (ensuring at least three distinct sources).
  \item \emph{Query Validation:} Refined queries are paired with their associated document snippets and evaluated against the ground truth. An internal retriever pipeline is used to simulate the reasoning path. A query is accepted only if the system can reconstruct the ground truth from the provided snippets, indicating sufficient evidence coverage and reasoning fidelity. 
\end{itemize}

\subsection{Dataset Statistics}
Using the MultiHop-RAG Corpus \citep{Tang2024MultiHopRAGBR}, we initially generated 1,173 questions and retained 674 high-quality queries after filtering. On average, each query spans 4.1 chunks and 3.4 documents (4 entities per question and 9 entities per answer). 

\paragraph{Sample Synthetic Query.} 

\begin{quote}
\textit{How has Manchester United's on-field performance under Erik ten Hag evolved amid the impending Ratcliffe ownership transition, considering their pressing statistics, Champions League failure, and potential managerial succession plans?}
\end{quote}

\textbf{Ground-truth answer}:  
\textit{"Manchester United's current situation reflects a complex intersection of tactical evolution and organizational transition. While Ten Hag has shown some statistical improvements in certain areas---notably with United ranking at the top of the Premier League in high ball regains and middle-third possession win..."}

This demands the system's ability to integrate performance metrics (pressing and possession statistics), organizational context (ownership transition), and multi-document references (Champions League details, managerial prospects).

\section{Experimental Setup}
\label{sec:experiment_setup}

We describe our experimental design, including datasets, indexing procedures, retrieval approaches, and evaluation metrics.

\subsection{Datasets}
\label{subsec:datasets}

Our experiments span five diverse datasets (Table~\ref{tab:datasets}), selected to evaluate multi-hop reasoning, domain-specific language, and tabular/context integration:

\textbf{MultiHop-RAG} \citep{Tang2024MultiHopRAGBR}. Multi-document QA dataset with queries that require synthesizing evidence from multiple sources, covering inference, temporal reasoning, comparisons, and null cases.

\textbf{SEC-10Q} \citep{docugami_kg_rag_2023}. Quarterly financial reports (Form 10-Q) from publicly traded companies, filed with the U.S. Securities and Exchange Commission. These reports include financial statements, management discussions, and disclosures of market risks.

\textbf{ConcurrentQA} \citep{concurrentqa2023}. A multi-hop QA benchmark requiring reasoning across both public (Wikipedia) and private (emails) data sources. It builds on HotpotQA \citep{yang2018hotpotqa} and evaluates the ability to handle complex cross-domain queries.

\textbf{NTSB} \citep{docugami_kg_rag_2023}. Corpus of aviation accident and incident reports from the National Transportation Safety Board. Each report details the event date, location, aircraft involved, and probable causal factors.

\textbf{WikiHowQA} \citep{2023-wikihowqa}. Derived from WikiHow articles, with community-generated questions, along with step-by-step procedures.

\begin{table*}[ht]
\centering
\caption{Overview of the datasets used for evaluation.}
\label{tab:datasets}    
\begin{tabular}{@{}lcccp{10cm}@{}}
\toprule
\textbf{Dataset} & \textbf{Domain} & \textbf{\# Queries} & \textbf{\# Docs} & \textbf{Sample Query} \\
\midrule
MultiHop-RAG     & News Articles    & 2,556  & 609     & Who is the individual associated with OpenAI, recognized for both his vision of AI agents and his generosity and has made headlines in both Fortune and TechCrunch for his controversial departure? \\
SEC-10Q          & Finance          & 195    & 20      & How has Apple's revenue from iPhone sales fluctuated across quarters? \\
NTSB             & Aviation         & 197    & 20      & Which operators have been involved in fatal accidents with amateur-built aircraft? \\
WikiHowQA        & General          & 300    & 5,000   & How to grow orchids in a greenhouse? \\
ConcurrentQA     & Email            & 400    & 13,500  & Which OpenTable.com investor also invested in Acta? \\
\bottomrule
\end{tabular}
\end{table*}

\subsection{Indexing Steps}
All datasets undergo a four-step indexing procedure. Steps 2--3 rely on an LLM (\texttt{Claude--3 Sonnet} \citep{anthropic_claude3_2024} used here) to relate domain-relevant entities and topics, with fine-grained propositions. We use AWS Neptune Analytics for the graph store and AWS OpenSearch for vector storage. At the time of writing, the framework also supports PostgreSQL with pgVector, FalkorDB, and MosaicML.

\begin{enumerate}
    \item \textbf{Chunking.} Each document is segmented into 300-token chunks, with a 20\% overlap to preserve context. 
    \item \textbf{Domain-Adaptive Refinement.} We sample five chunks from each dataset to infer domain-specific entity types and topics (e.g., financial terminology in SEC-10Q). This ensures that HLG reflects relevant domain concepts.
    \item \textbf{Proposition \& Graph Construction.} From each chunk, we extract fine-grained propositions, and then link them with topics, entities, and relationships in HLG.
    \item \textbf{Embedding Generation.} Embeddings for topics, statements, or both, per retrieval method. We use \emph{Cohere-Embed-English-v3} to compute 1024-dimensional vectors \cite{reimers2023embedv3}.
\end{enumerate}

\subsection{Retrieval Approaches}

We evaluate three baseline methods and four HLG-based retrieval approaches. For all methods, the context window is fixed at 10 chunks, which corresponds to approximately 3,000 tokens. Following retrieval, (\texttt{BAAI/bge-reranker-v2-minicpm-layerwise} \citep{li2023reranker}) selects the final set of passages (chunks/statements). This reranker, based on a streamlined 2B-LLM, processes queries and retrieved statements jointly, enabling deeper context understanding compared to embedding-based methods that rely on fixed vector representations. All methods, including baselines, use the same LLM (\texttt{Claude-3 Sonnet} \cite{anthropic_claude3_2024}) and identical prompting for answer generation. We avoid techniques like Chain-of-Thought (CoT) prompting, as they can conflate retrieval effectiveness with reasoning ability. Our focus is strictly on retrieval augmentation. Importantly, \texttt{Claude-3 Sonnet}'s training data was frozen in August 2023 \citep{anthropic_claude3_2024}, predating the MultiHop-RAG articles (Sept-Dec 2023) \citep{Tang2024MultiHopRAGBR}. Thus, its performance on this dataset cannot be attributed to memorization.

\subsubsection{Baselines}
\label{subsec:baselines}
\begin{itemize}
    \item \textbf{Naive RAG} (B0). Retrieves 10 chunks via VSS. VSS uses a dense retriever, applying k-nearest neighbour (kNN) algorithms to rank chunks based on the similarity between the query and chunk embedding \citep{cover1967knn}.
    \item \textbf{Naive RAG (w/ reranking)} (B1). Retrieves 50 chunks initially, applies the reranking model, and prunes the output to retain the top 10 chunks for the final set.
    \item \textbf{Entity-Linking Expansion} (E1). A simpler graph-like baseline. Retrieves 5 chunks via VSS, expands with one-hop entity links from HLG, merges up to 50 nodes (chunks associated with linked entities), then reranks to top 10 chunks. This is similar to the triplet extraction idea in some related works.
\end{itemize}

\subsubsection{Our Approaches}
\begin{itemize}
    \item \textbf{StatementRAG} (SRAG). Performs VSS on individual statements (rather than entire chunks). Retrieve 100 statements, then reranks and truncates the top results.
    \item \textbf{StatementGraphRAG} (SGRAG). Extends SRAG with graph-based expansions. Starting from 100 initial statements, we run a beam search over graph neighbours (width=50, depth=3). The expanded pool is reranked and truncated.
    \item \textbf{TopicRAG} (TRAG). Retrieves 50 topics based on VSS. The associated statements for each topic are retrieved. The statements are then reranked and truncated.
    \item \textbf{TopicGraphRAG} (TGRAG). Combines topic-level retrieval with graph expansions. First, retrieves 50 topics via VSS, expands using beam search (width=50, depth=3), and reranks.
\end{itemize}

\subsection{Chunks as the Generation Unit}
\label{subsec:chunksgenerationunit}
In certain domains, compliance regulations mandate the use of original text blocks rather than LLM-generated statements. To address this, we evaluate two HLG methods (SGRAG and TGRAG) in a "chunk-based" variant. This approach leverages the detailed connections between individual statements during graph expansion, while ensuring the generation prompt is composed solely of original text blocks. In the final step, the top-ranked statements are traced back to their source chunks. We select up to 10 unique chunks and apply a diversity filter to remove near-duplicates, further promoting chunk-level diversity in the final output.

\subsection{Evaluation Metrics}
\label{subsec:eval_metrics}
\textbf{Correctness} is used for single-answer datasets (MultiHop-RAG, SEC-10Q, ConcurrentQA) and \textbf{Answer Recall} for multi-answer datasets (NTSB, WikiHowQA). \emph{Correctness} assesses if the generated answer semantically contains all necessary information from the gold answer, even if phrasing differs. For single-answer datasets, correctness is akin to accuracy, requiring a semantically correct response (e.g., a "true" response may be conceptually correct for a "yes" ground truth, even if not an exact string match). This is preferred over exact match, which can be overly rigid for free-form, multi-part responses. \emph{Answer Recall} measures the proportion of gold answer facts correctly included in the generated answer. For instance, listing 4 of 5 steps scores 0\% on strict correctness but 80\% on recall. This captures partial coverage. For the synthetic dataset, we evaluate the retrievers using claim recall (CR) and context precision (CP) from RAGChecker \citep{ru2024ragchecker}.

\section{Results}
\label{sec:results}
We present retrieval evaluations across multiple datasets and metrics. We compare (a) chunk- vs.\ statement-level retrieval, (b) correctness vs.\ recall, and (c) pairwise comparisons against baselines.

\subsection{Overall Retrieval Performance}
Table \ref{tab:combined_results} reports results on three datasets with a single gold answer (MultiHop-RAG, SEC-10Q, and ConcurrentQA) and two datasets allowing multiple valid answers (NTSB, WikiHowQA).

\begin{table*}[ht]
\centering
\caption{Correctness and answer recall results. \textbf{Bold} = highest in column; \underline{underlined} = second highest.}
\label{tab:combined_results}
\renewcommand{\arraystretch}{1.2}
\setlength{\tabcolsep}{4pt}      
\small
\begin{tabular}{lccccccc}
\toprule
\textbf{} 
& \multicolumn{3}{c}{\textbf{Correctness}} 
& \multirow{2}{*}{\shortstack{\textbf{Correctness} \\ \textbf{Average}}} 
& \multicolumn{2}{c}{\textbf{Answer Recall}} 
& \multirow{2}{*}{\shortstack{\textbf{Answer Recall} \\ \textbf{Average}}} \\
\cmidrule(lr){2-4} \cmidrule(lr){6-7}
& \textbf{MultiHop-RAG} 
& \textbf{SEC-10Q} 
& \textbf{ConcurrentQA} 
& 
& \textbf{NTSB} 
& \textbf{WikiHowQA} 
& \\
\midrule
\textit{B0 (Naive RAG)} 
 & 78.7\%
 & 64.1\% 
 & 43.3\% 
 & 62.0\%
 & 24.2\% 
 & \underline{69.6\%} 
 & 46.9\% \\
\textit{B1 (Naive RAG + rerank)} 
 & 82.8\%
 & 64.6\% 
 & 51.0\% 
 & 66.1\%
 & 31.1\% 
 & \textbf{70.5\%} 
 & 50.8\% \\
 \textit{E1 (Entity-Linking RAG)}
 & 82.6\%
 & 61.5\%
 & 52.8\%
 & 65.6\%
 & 31.2\%
 & 70.0\%
 & 50.6\% \\
\midrule
\textit{SRAG} 
 & \textbf{87.0\%}
 & 69.2\%
 & 51.3\% 
 & 69.2\%
 & 33.4\% 
 & 65.0\%
 & 49.2\%\\
\textit{SGRAG} 
 & \underline{86.9\%}
 & \underline{73.9\%} 
 & 57.3\%
 & \underline{72.7\%}
 & 34.8\% 
 & 67.1\% 
 & 50.9\%\\
\textit{SGRAG-0.5\%} 
 & \underline{86.9\%}
 & \textbf{74.4\%}
 & \underline{59.5\%}
 & \textbf{73.6\%}
 & \underline{36.1\%} 
 & 68.7\% 
 & \underline{52.4\%} \\
\midrule
\textit{TRAG} 
 & 84.2\%
 & 70.3\%
 & 53.8\% 
 & 69.4\%
 & 35.7\% 
 & 68.3\% 
 & 52.0\% \\
\textit{TGRAG} 
 & 84.5\%
 & 72.3\%
 & \textbf{59.8\%} 
 & 72.2\%
 & \textbf{39.3\%}
 & 68.2\% 
 &\textbf{ 53.8\%}
 \\
\bottomrule
\end{tabular}
\end{table*}

\noindent
\textbf{Discussion.}
Among the baselines, B1 achieves 66.1\% correctness averaged over the three single-answer datasets, surpassing B0 by 4.1 absolute percentage points. E1, which performs simple entity linking, shows minimal improvement over B1 on average (65.6\% correctness) and underperforms on SEC-10Q, suggesting that naive entity expansion alone is not consistently beneficial. In contrast, every HLG-based method (SRAG, SGRAG, TRAG, TGRAG, and their variants) outperforms B1 in at least one dimension (correctness or recall). SGRAG-0.5\% attains the highest average correctness (73.6\%), an overall improvement of 7.5 points over B1. This method also improves average answer recall (52.4\%), up from 50.8\% with B1.

On multi-answer tasks (NTSB and WikiHowQA), TGRAG yields the best average recall (53.8\%), outperforming SGRAG-0.5\% by 1.4 absolute points. Meanwhile, statement-based methods such as SRAG and SGRAG demonstrate consistently higher correctness on single-answer datasets, with SRAG reaching 87.0\% on MultiHop-RAG and SGRAG-0.5\% yielding 74.4\% on SEC-10Q. These results suggest that statement-level retrieval is adept at pinpointing a single correct piece of evidence, whereas topic-based retrieval may capture a broader set of relevant facts.
The WikiHowQA dataset predominantly comprises single-hop, procedural questions (further discussed in Appendix \ref{subsubsec:single-hop}). In such cases, multi-hop graph expansions (as in SGRAG/TGRAG) can introduce noise or irrelevant hops, potentially degrading performance compared to simpler methods like B1, which performs best on this dataset. While HLG-based methods underperform slightly on WikiHowQA, they consistently outperform baselines on all true multi-hop datasets (MultiHop-RAG, ConcurrentQA). This dataset-specific behaviour shows that HLG is optimized for multi-hop retrieval, and single-hop cases may benefit from reduced complexity or early-stopping heuristics.

\subsection{HLG Chunk-Based Variants}
Because some applications require preserving the original text at the chunk level, we also evaluate Chunk-SGRAG and Chunk-TGRAG. These methods internally retrieve statements using graph traversal, but for the final output, they map the selected top statements back to their parent chunks. Table \ref{tab:chunk} shows how these chunk-level variants compare to the baselines.

\begin{table*}[ht]
\centering
\caption{Chunk-based variants of StatementGraphRAG and TopicGraphRAG.}
\label{tab:chunk}
\renewcommand{\arraystretch}{1.2} 
\setlength{\tabcolsep}{4pt}      
\small 
\begin{tabular}{lccccccc}
\toprule
\textbf{} 
& \multicolumn{3}{c}{\textbf{Correctness}} 
& \multirow{2}{*}{\shortstack{\textbf{Correctness} \\ \textbf{Average}}} 
& \multicolumn{2}{c}{\textbf{Answer Recall}} 
& \multirow{2}{*}{\shortstack{\textbf{Answer Recall} \\ \textbf{Average}}} \\
\cmidrule(lr){2-4} \cmidrule(lr){6-7}
& \textbf{MultiHop-RAG} 
& \textbf{SEC-10Q} 
& \textbf{ConcurrentQA} 
& 
& \textbf{NTSB} 
& \textbf{WikiHowQA} 
& \\
\midrule
\textit{Chunk-SGRAG}  
 & 86.1\% 
 & 65.6\% 
 & 66.8\% 
 & 72.8\% 
 & 25.8\% 
 & 67.2\% 
 & 46.5\% \\
\textit{Chunk-SGRAG-0.5\%}  
 & 86.1\% 
 & 66.2\% 
 & 67.0\% 
 & 73.1\% 
 & 25.8\% 
 & 67.7\% 
 & 46.7\% \\
\midrule
\textit{Chunk-TGRAG}  
 & 86.3\% 
 & 72.8\% 
 & 61.8\% 
 & 73.6\% 
 & 27.7\% 
 & 67.0\% 
 & 47.3\% \\
\textit{Chunk-TGRAG-0.5\%}  
 & 86.0\% 
 & 73.3\% 
 & 64.3\% 
 & 74.5\% 
 & 26.7\% 
 & 67.5\% 
 & 47.1\% \\
\bottomrule
\end{tabular}
\end{table*}

\noindent
\textbf{Discussion.} Relative to naive chunk-only retrieval (B0 or B1), the hybrid chunk-based variants consistently yield higher correctness. On ConcurrentQA, for example, Chunk-SGRAG-0.5\% reaches 67.0\% correctness, a 23.7-point increase over B0 and 16.0 points over B1. Despite initially retrieving content at the finer statement granularity, the final output remains chunked. This suggests that graph-based expansions are beneficial even if the generation context window must be text segments (chunks).

\subsection{Synthetic Dataset Evaluation}
\subsubsection{RAGChecker Evaluation}
We further investigated retrieval quality on our synthetic MultiHop-RAG subset using RAGChecker \citep{ru2024ragchecker}, which provides fine-grained retriever metrics such as claim recall (CR) and context precision (CP). Table \ref{tab:synthetic_eval} reports these metrics for the baselines (B0, B1) and our primary HLG-based methods (SGRAG, TGRAG). Generator metrics are omitted because all systems use the same LLM for answer generation.

\begin{table}[ht]
\centering
\caption{RAGChecker metrics on a synthetic MultiHop-RAG subset. Retriever is assessed using claim recall (CR) and context precision (CP).}
\label{tab:synthetic_eval}
\renewcommand{\arraystretch}{1.1} 
\setlength{\tabcolsep}{6pt}      
\small
\begin{tabular}{lccccc}
\toprule
\textbf{} 
& \multicolumn{3}{c}{\textbf{Overall}} 
& \multicolumn{2}{c}{\textbf{Retriever}} \\
\cmidrule(lr){2-4} \cmidrule(lr){5-6}
& \textbf{Precision (↑)} 
& \textbf{Recall (↑)} 
& \textbf{F1 (↑)} 
& \textbf{CR (↑)} 
& \textbf{CP (↑)} \\
\midrule
\textit{B0} 
 & \textbf{51.2} 
 & 46.0 
 & 46.1 
 & 58.3 
 & 78.8 \\
\textit{B1} 
 & 51.1 
 & 46.3 
 & 46.3 
 & 62.2 
 & \textbf{81.5} \\
 \midrule
\textit{SGRAG} 
 & 50.8
 & 48.6 
 & 47.5 
 & 59.5 
 & 50.9 \\
 \midrule
\textit{TGRAG} 
 & 50.8
 & \textbf{49.3} 
 & \textbf{47.9} 
 & \textbf{67.6} 
 & 66.1 \\
\bottomrule
\end{tabular}
\end{table}

\noindent
\textbf{Discussion.} Although the baselines (B0, B1) show slightly higher context precision (78.8\%-81.5\%), both SGRAG and TGRAG surpass them in recall and F1. In particular, TGRAG achieves 49.3\% recall and 47.9\% F1, exceeding B1 by 3.0 and 1.6 absolute points, respectively. The stronger claim recall (67.6\% vs.\ 62.2\% for B1) also showcases TGRAG’s multi-hop capabilities, as it is more likely to gather all relevant facts for each query.

\subsubsection{Pairwise Evaluation: Win Rates}
We also conducted a head-to-head comparison of SGRAG and TGRAG against the baselines. In each comparison, an LLM evaluator was shown the two retrieved answers in random order to mitigate any position bias. Table~\ref{tab:win_rates} presents the percentage of times each method won, lost, or tied.

\begin{table}[ht]
\centering
\caption{Pairwise comparisons of StatementGraphRAG and TopicGraphRAG vs. chunk-based baselines (B0 and B1).}
\label{tab:win_rates}
\renewcommand{\arraystretch}{1.1}
\setlength{\tabcolsep}{4pt}     
\small
\begin{tabular}{lccccc}
\toprule
\textbf{} & \textbf{Baseline} & \textbf{Win (\%)} & \textbf{Loss (\%)} & \textbf{Tie (\%)} & \textbf{Win/Loss Ratio} \\
\midrule
\multirow{2}{*}{\textit{SGRAG}} 
 & B0 & 74.8 & 21.5 & 3.7 & 3.5 \\
 & B1 & 74.8 & 22.9 & 2.8 & 3.3 \\
\midrule
\multirow{2}{*}{\textit{TGRAG}} 
 & B0 & 78.3 & 17.4 & 4.3 & 4.5 \\
 & B1 & 76.3 & 18.4 & 5.3 & 4.2 \\
\bottomrule
\end{tabular}
\end{table}

\noindent
\textbf{Discussion.} Both SGRAG and TGRAG significantly outperform the chunk-based baselines, winning over 74\% of head-to-head comparisons. TGRAG achieves a slightly higher win rate against B0 (78.3\%) compared to SGRAG (74.8\%), reflecting its stronger coverage of multiple answers. These consistent pairwise improvements highlight that graph-based retrieval methods (whether statement- or topic-centered) tend to produce more relevant and higher-quality evidence than VSS retrieval.

\subsection{Summary of Findings}
For single-answer tasks, statement-based retrieval methods (SRAG, SGRAG) consistently achieve higher correctness, averaging 72.7\%-73.6\%, compared to 62.0\%-66.1\% for naive baselines. In contrast, topic-based approaches (TRAG, TGRAG) perform best on multi-answer tasks. Graph expansions yield significant improvements: SGRAG-0.5\% exceeds the naive baseline (B0) by 8-12 points in correctness across single-answer datasets, while TGRAG improves answer recall by more than 10 points over B0 in multi-answer settings. Introducing a modest diversity threshold (0.5\%) further enhances coverage by filtering near-duplicate statements, resulting in an additional 1-point gain in correctness. Even when chunk-based outputs are used, leveraging statement-level graph traversals internally enables retrieval of more relevant content, demonstrating that fine-grained graph expansions enhance overall chunk selection.

\section{Conclusion}
\label{sec:conclusion}
We presented a Hierarchical Lexical Graph framework that supports fine-grained, multi-hop retrieval for QA tasks. Our experiments demonstrated that statement-level and topic-level retrieval consistently outperform baselines across correctness, answer recall, and pairwise judgements, with TopicGraphRAG especially strong in multi-hop scenarios demanding broader coverage. The graph expansion strategies capture inter-document relationships more effectively, while a lightweight reranking model balances performance with reduced latency. Despite higher indexing costs and domain constraints, experimental results, including strong statement-to-source alignment, show viability for structured and unstructured data. Future work will focus on refining proposition extraction with smaller models, incorporating adaptive multi-hop detection, and expanding domain specialization, thereby advancing the accuracy and efficiency of multi-hop RAG systems in real-world applications.

\medskip
{
\small
\bibliographystyle{ACM-Reference-Format}
\bibliography{references}


\begin{thebibliography}{23}


\ifx \showCODEN    \undefined \def \showCODEN     #1{\unskip}     \fi
\ifx \showDOI      \undefined \def \showDOI       #1{#1}\fi
\ifx \showISBNx    \undefined \def \showISBNx     #1{\unskip}     \fi
\ifx \showISBNxiii \undefined \def \showISBNxiii  #1{\unskip}     \fi
\ifx \showISSN     \undefined \def \showISSN      #1{\unskip}     \fi
\ifx \showLCCN     \undefined \def \showLCCN      #1{\unskip}     \fi
\ifx \shownote     \undefined \def \shownote      #1{#1}          \fi
\ifx \showarticletitle \undefined \def \showarticletitle #1{#1}   \fi
\ifx \showURL      \undefined \def \showURL       {\relax}        \fi
\providecommand\bibfield[2]{#2}
\providecommand\bibinfo[2]{#2}
\providecommand\natexlab[1]{#1}
\providecommand\showeprint[2][]{arXiv:#2}

\bibitem[Anthropic(2024a)]%
        {anthropic_claude3_2024}
\bibfield{author}{\bibinfo{person}{Anthropic}.} \bibinfo{year}{2024}\natexlab{a}.
\newblock \bibinfo{title}{Claude 3 Model Card}.
\newblock
\newblock
\urldef\tempurl%
\url{https://www.anthropic.com/model_cards/claude_3.pdf}
\showURL{%
\tempurl}


\bibitem[Anthropic(2024b)]%
        {anthropic2024claude35}
\bibfield{author}{\bibinfo{person}{Anthropic}.} \bibinfo{year}{2024}\natexlab{b}.
\newblock \bibinfo{title}{Model Card Addendum: Claude 3.5 Haiku and Upgraded Claude 3.5 Sonnet}.
\newblock
\newblock
\urldef\tempurl%
\url{https://assets.anthropic.com/m/1cd9d098ac3e6467/original/Claude-3-Model-Card-October-Addendum.pdf}
\showURL{%
\tempurl}


\bibitem[Arora et~al\mbox{.}(2023)]%
        {concurrentqa2023}
\bibfield{author}{\bibinfo{person}{Simran Arora}, \bibinfo{person}{Patrick Lewis}, \bibinfo{person}{Angela Fan}, \bibinfo{person}{Jacob Kahn}, {and} \bibinfo{person}{Christopher R{\'e}}.} \bibinfo{year}{2023}\natexlab{}.
\newblock \showarticletitle{Reasoning over Public and Private Data in Retrieval-Based Systems}.
\newblock \bibinfo{journal}{\emph{Transactions of the Association for Computational Linguistics}}  \bibinfo{volume}{11} (\bibinfo{year}{2023}), \bibinfo{pages}{902--921}.
\newblock
\urldef\tempurl%
\url{https://doi.org/10.1162/tacl_a_00580}
\showDOI{\tempurl}


\bibitem[Bahdanau et~al\mbox{.}(2016)]%
        {bahdanau2016neuralmachinetranslationjointly}
\bibfield{author}{\bibinfo{person}{Dzmitry Bahdanau}, \bibinfo{person}{Kyunghyun Cho}, {and} \bibinfo{person}{Yoshua Bengio}.} \bibinfo{year}{2016}\natexlab{}.
\newblock \bibinfo{title}{Neural Machine Translation by Jointly Learning to Align and Translate}.
\newblock
\newblock
\showeprint[arxiv]{1409.0473}~[cs.CL]
\urldef\tempurl%
\url{https://arxiv.org/abs/1409.0473}
\showURL{%
\tempurl}


\bibitem[Besta et~al\mbox{.}(2024)]%
        {besta2024multihead}
\bibfield{author}{\bibinfo{person}{Maciej Besta}, \bibinfo{person}{Ales Kubicek}, \bibinfo{person}{Roman Niggli}, \bibinfo{person}{Robert Gerstenberger}, \bibinfo{person}{Lucas Weitzendorf}, \bibinfo{person}{Mingyuan Chi}, \bibinfo{person}{Patrick Iff}, \bibinfo{person}{Joanna Gajda}, \bibinfo{person}{Piotr Nyczyk}, \bibinfo{person}{Jürgen Müller}, \bibinfo{person}{Hubert Niewiadomski}, \bibinfo{person}{Marcin Chrapek}, \bibinfo{person}{Michał Podstawski}, {and} \bibinfo{person}{Torsten Hoefler}.} \bibinfo{year}{2024}\natexlab{}.
\newblock \bibinfo{title}{Multi-Head RAG: Solving Multi-Aspect Problems with LLMs}.
\newblock
\newblock
\showeprint[arxiv]{2406.05085}~[cs.CL]
\urldef\tempurl%
\url{https://arxiv.org/abs/2406.05085}
\showURL{%
\tempurl}


\bibitem[Bolotova-Baranova et~al\mbox{.}(2023)]%
        {2023-wikihowqa}
\bibfield{author}{\bibinfo{person}{Valeriia Bolotova-Baranova}, \bibinfo{person}{Vladislav Blinov}, \bibinfo{person}{Sofya Filippova}, \bibinfo{person}{Falk Scholer}, {and} \bibinfo{person}{Mark Sanderson}.} \bibinfo{year}{2023}\natexlab{}.
\newblock \showarticletitle{{W}iki{H}ow{QA}: A Comprehensive Benchmark for Multi-Document Non-Factoid Question Answering}. In \bibinfo{booktitle}{\emph{Proceedings of the 61st Annual Meeting of the Association for Computational Linguistics (Volume 1: Long Papers)}}, \bibfield{editor}{\bibinfo{person}{Anna Rogers}, \bibinfo{person}{Jordan Boyd-Graber}, {and} \bibinfo{person}{Naoaki Okazaki}} (Eds.). \bibinfo{publisher}{Association for Computational Linguistics}, \bibinfo{address}{Toronto, Canada}, \bibinfo{pages}{5291--5314}.
\newblock
\urldef\tempurl%
\url{https://doi.org/10.18653/v1/2023.acl-long.290}
\showDOI{\tempurl}


\bibitem[Chen et~al\mbox{.}(2024)]%
        {chen2024dense}
\bibfield{author}{\bibinfo{person}{Tong Chen}, \bibinfo{person}{Hongwei Wang}, \bibinfo{person}{Sihao Chen}, \bibinfo{person}{Wenhao Yu}, \bibinfo{person}{Kaixin Ma}, \bibinfo{person}{Xinran Zhao}, \bibinfo{person}{Hongming Zhang}, {and} \bibinfo{person}{Dong Yu}.} \bibinfo{year}{2024}\natexlab{}.
\newblock \showarticletitle{Dense {X} Retrieval: What Retrieval Granularity Should We Use?}. In \bibinfo{booktitle}{\emph{Proceedings of the 2024 Conference on Empirical Methods in Natural Language Processing}}, \bibfield{editor}{\bibinfo{person}{Yaser Al-Onaizan}, \bibinfo{person}{Mohit Bansal}, {and} \bibinfo{person}{Yun-Nung Chen}} (Eds.). \bibinfo{publisher}{Association for Computational Linguistics}, \bibinfo{address}{Miami, Florida, USA}, \bibinfo{pages}{15159--15177}.
\newblock
\urldef\tempurl%
\url{https://doi.org/10.18653/v1/2024.emnlp-main.845}
\showDOI{\tempurl}


\bibitem[Cohere(2023)]%
        {reimers2023embedv3}
\bibfield{author}{\bibinfo{person}{Cohere}.} \bibinfo{year}{2023}\natexlab{}.
\newblock \bibinfo{title}{Introducing Embed v3}.
\newblock \bibinfo{howpublished}{\url{https://cohere.com/blog/introducing-embed-v3}}.
\newblock
\newblock
\shownote{Accessed: 2025-05-27}.


\bibitem[Cover and Hart(1967)]%
        {cover1967knn}
\bibfield{author}{\bibinfo{person}{T. Cover} {and} \bibinfo{person}{P. Hart}.} \bibinfo{year}{1967}\natexlab{}.
\newblock \showarticletitle{Nearest neighbor pattern classification}.
\newblock \bibinfo{journal}{\emph{IEEE Transactions on Information Theory}} \bibinfo{volume}{13}, \bibinfo{number}{1} (\bibinfo{year}{1967}), \bibinfo{pages}{21--27}.
\newblock
\urldef\tempurl%
\url{https://doi.org/10.1109/TIT.1967.1053964}
\showDOI{\tempurl}


\bibitem[De~Cao et~al\mbox{.}(2019)]%
        {gcnn}
\bibfield{author}{\bibinfo{person}{Nicola De~Cao}, \bibinfo{person}{Wilker Aziz}, {and} \bibinfo{person}{Ivan Titov}.} \bibinfo{year}{2019}\natexlab{}.
\newblock \showarticletitle{Question Answering by Reasoning Across Documents with Graph Convolutional Networks}. In \bibinfo{booktitle}{\emph{Proceedings of the 2019 Conference of the North {A}merican Chapter of the Association for Computational Linguistics: Human Language Technologies, Volume 1 (Long and Short Papers)}}, \bibfield{editor}{\bibinfo{person}{Jill Burstein}, \bibinfo{person}{Christy Doran}, {and} \bibinfo{person}{Thamar Solorio}} (Eds.). \bibinfo{publisher}{Association for Computational Linguistics}, \bibinfo{address}{Minneapolis, Minnesota}, \bibinfo{pages}{2306--2317}.
\newblock
\urldef\tempurl%
\url{https://doi.org/10.18653/v1/N19-1240}
\showDOI{\tempurl}


\bibitem[Edge et~al\mbox{.}(2024)]%
        {edge2024localglobalgraphrag}
\bibfield{author}{\bibinfo{person}{Darren Edge}, \bibinfo{person}{Ha Trinh}, \bibinfo{person}{Newman Cheng}, \bibinfo{person}{Joshua Bradley}, \bibinfo{person}{Alex Chao}, \bibinfo{person}{Apurva Mody}, \bibinfo{person}{Steven Truitt}, {and} \bibinfo{person}{Jonathan Larson}.} \bibinfo{year}{2024}\natexlab{}.
\newblock \bibinfo{title}{From Local to Global: A Graph RAG Approach to Query-Focused Summarization}.
\newblock
\newblock
\showeprint[arxiv]{2404.16130}~[cs.CL]
\urldef\tempurl%
\url{https://arxiv.org/abs/2404.16130}
\showURL{%
\tempurl}


\bibitem[Fang et~al\mbox{.}(2020)]%
        {Fang2019HierarchicalGN}
\bibfield{author}{\bibinfo{person}{Yuwei Fang}, \bibinfo{person}{Siqi Sun}, \bibinfo{person}{Zhe Gan}, \bibinfo{person}{Rohit Pillai}, \bibinfo{person}{Shuohang Wang}, {and} \bibinfo{person}{Jingjing Liu}.} \bibinfo{year}{2020}\natexlab{}.
\newblock \showarticletitle{Hierarchical Graph Network for Multi-hop Question Answering}. In \bibinfo{booktitle}{\emph{Proceedings of the 2020 Conference on Empirical Methods in Natural Language Processing (EMNLP)}}, \bibfield{editor}{\bibinfo{person}{Bonnie Webber}, \bibinfo{person}{Trevor Cohn}, \bibinfo{person}{Yulan He}, {and} \bibinfo{person}{Yang Liu}} (Eds.). \bibinfo{publisher}{Association for Computational Linguistics}, \bibinfo{address}{Online}, \bibinfo{pages}{8823--8838}.
\newblock
\urldef\tempurl%
\url{https://doi.org/10.18653/v1/2020.emnlp-main.710}
\showDOI{\tempurl}


\bibitem[Gao et~al\mbox{.}(2023)]%
        {Gao2023RetrievalAugmentedGF}
\bibfield{author}{\bibinfo{person}{Yunfan Gao}, \bibinfo{person}{Yun Xiong}, \bibinfo{person}{Xinyu Gao}, \bibinfo{person}{Kangxiang Jia}, \bibinfo{person}{Jinliu Pan}, \bibinfo{person}{Yuxi Bi}, \bibinfo{person}{Yi Dai}, \bibinfo{person}{Jiawei Sun}, \bibinfo{person}{Qianyu Guo}, \bibinfo{person}{Meng Wang}, {and} \bibinfo{person}{Haofen Wang}.} \bibinfo{year}{2023}\natexlab{}.
\newblock \showarticletitle{Retrieval-Augmented Generation for Large Language Models: {A} Survey}.
\newblock \bibinfo{journal}{\emph{CoRR}}  \bibinfo{volume}{abs/2312.10997} (\bibinfo{year}{2023}).
\newblock
\urldef\tempurl%
\url{https://doi.org/10.48550/ARXIV.2312.10997}
\showDOI{\tempurl}
\showeprint[arXiv]{2312.10997}


\bibitem[He et~al\mbox{.}(2023)]%
        {graphAttention}
\bibfield{author}{\bibinfo{person}{Yunjie He}, \bibinfo{person}{Philip~John Gorinski}, \bibinfo{person}{Ieva Staliunaite}, {and} \bibinfo{person}{Pontus Stenetorp}.} \bibinfo{year}{2023}\natexlab{}.
\newblock \bibinfo{title}{Graph Attention with Hierarchies for Multi-hop Question Answering}.
\newblock
\newblock
\showeprint[arxiv]{2301.11792}~[cs.CL]
\urldef\tempurl%
\url{https://arxiv.org/abs/2301.11792}
\showURL{%
\tempurl}


\bibitem[Jaffri(2023)]%
        {docugami_kg_rag_2023}
\bibfield{author}{\bibinfo{person}{Taqi Jaffri}.} \bibinfo{year}{2023}\natexlab{}.
\newblock \bibinfo{title}{Announcing Docugami Knowledge Graph Retrieval Augmented Generation (KG-RAG) Datasets in the LlamaHub}.
\newblock
\newblock
\urldef\tempurl%
\url{https://www.docugami.com/blog/kg-rag-datasets-llama-index}
\showURL{%
\tempurl}


\bibitem[Kim et~al\mbox{.}(2024)]%
        {Kim2024Prometheus2A}
\bibfield{author}{\bibinfo{person}{Seungone Kim}, \bibinfo{person}{Juyoung Suk}, \bibinfo{person}{Shayne Longpre}, \bibinfo{person}{Bill~Yuchen Lin}, \bibinfo{person}{Jamin Shin}, \bibinfo{person}{Sean Welleck}, \bibinfo{person}{Graham Neubig}, \bibinfo{person}{Moontae Lee}, \bibinfo{person}{Kyungjae Lee}, {and} \bibinfo{person}{Minjoon Seo}.} \bibinfo{year}{2024}\natexlab{}.
\newblock \showarticletitle{Prometheus 2: An Open Source Language Model Specialized in Evaluating Other Language Models}. In \bibinfo{booktitle}{\emph{Proceedings of the 2024 Conference on Empirical Methods in Natural Language Processing}}, \bibfield{editor}{\bibinfo{person}{Yaser Al-Onaizan}, \bibinfo{person}{Mohit Bansal}, {and} \bibinfo{person}{Yun-Nung Chen}} (Eds.). \bibinfo{publisher}{Association for Computational Linguistics}, \bibinfo{address}{Miami, Florida, USA}, \bibinfo{pages}{4334--4353}.
\newblock
\urldef\tempurl%
\url{https://doi.org/10.18653/v1/2024.emnlp-main.248}
\showDOI{\tempurl}


\bibitem[Lewis et~al\mbox{.}(2020)]%
        {lewis2021rag}
\bibfield{author}{\bibinfo{person}{Patrick Lewis}, \bibinfo{person}{Ethan Perez}, \bibinfo{person}{Aleksandra Piktus}, \bibinfo{person}{Fabio Petroni}, \bibinfo{person}{Vladimir Karpukhin}, \bibinfo{person}{Naman Goyal}, \bibinfo{person}{Heinrich K\"{u}ttler}, \bibinfo{person}{Mike Lewis}, \bibinfo{person}{Wen-tau Yih}, \bibinfo{person}{Tim Rockt\"{a}schel}, \bibinfo{person}{Sebastian Riedel}, {and} \bibinfo{person}{Douwe Kiela}.} \bibinfo{year}{2020}\natexlab{}.
\newblock \showarticletitle{Retrieval-augmented generation for knowledge-intensive NLP tasks}. In \bibinfo{booktitle}{\emph{Proceedings of the 34th International Conference on Neural Information Processing Systems}} (Vancouver, BC, Canada) \emph{(\bibinfo{series}{NIPS '20})}. \bibinfo{publisher}{Curran Associates Inc.}, \bibinfo{address}{Red Hook, NY, USA}, Article \bibinfo{articleno}{793}, \bibinfo{numpages}{16}~pages.
\newblock
\showISBNx{9781713829546}


\bibitem[Li et~al\mbox{.}(2023)]%
        {li2023reranker}
\bibfield{author}{\bibinfo{person}{Chaofan Li}, \bibinfo{person}{Zheng Liu}, \bibinfo{person}{Shitao Xiao}, {and} \bibinfo{person}{Yingxia Shao}.} \bibinfo{year}{2023}\natexlab{}.
\newblock \showarticletitle{Making Large Language Models A Better Foundation For Dense Retrieval}.
\newblock \bibinfo{journal}{\emph{CoRR}}  \bibinfo{volume}{abs/2312.15503} (\bibinfo{year}{2023}).
\newblock
\urldef\tempurl%
\url{https://doi.org/10.48550/arXiv.2312.15503}
\showURL{%
\tempurl}


\bibitem[Ru et~al\mbox{.}(2024)]%
        {ru2024ragchecker}
\bibfield{author}{\bibinfo{person}{Dongyu Ru}, \bibinfo{person}{Lin Qiu}, \bibinfo{person}{Xiangkun Hu}, \bibinfo{person}{Tianhang Zhang}, \bibinfo{person}{Peng Shi}, \bibinfo{person}{Shuaichen Chang}, \bibinfo{person}{Cheng Jiayang}, \bibinfo{person}{Cunxiang Wang}, \bibinfo{person}{Shichao Sun}, \bibinfo{person}{Huanyu Li}, \bibinfo{person}{Zizhao Zhang}, \bibinfo{person}{Binjie Wang}, \bibinfo{person}{Jiarong Jiang}, \bibinfo{person}{Tong He}, \bibinfo{person}{Zhiguo Wang}, \bibinfo{person}{Pengfei Liu}, \bibinfo{person}{Yue Zhang}, {and} \bibinfo{person}{Zheng Zhang}.} \bibinfo{year}{2024}\natexlab{}.
\newblock \showarticletitle{RAGChecker: A Fine-grained Framework for Diagnosing Retrieval-Augmented Generation}. In \bibinfo{booktitle}{\emph{Advances in Neural Information Processing Systems}}, \bibfield{editor}{\bibinfo{person}{A.~Globerson}, \bibinfo{person}{L.~Mackey}, \bibinfo{person}{D.~Belgrave}, \bibinfo{person}{A.~Fan}, \bibinfo{person}{U.~Paquet}, \bibinfo{person}{J.~Tomczak}, {and} \bibinfo{person}{C.~Zhang}} (Eds.), Vol.~\bibinfo{volume}{37}. \bibinfo{publisher}{Curran Associates, Inc.}, \bibinfo{pages}{21999--22027}.
\newblock
\urldef\tempurl%
\url{https://proceedings.neurips.cc/paper_files/paper/2024/file/27245589131d17368cccdfa990cbf16e-Paper-Datasets_and_Benchmarks_Track.pdf}
\showURL{%
\tempurl}


\bibitem[Sparck~Jones(1972)]%
        {tfidf_1972}
\bibfield{author}{\bibinfo{person}{K. Sparck~Jones}.} \bibinfo{year}{1972}\natexlab{}.
\newblock \showarticletitle{A Statistical Interpretation of Term Specificity and Its Application in Retrieval}.
\newblock \bibinfo{journal}{\emph{Journal of Documentation}} \bibinfo{volume}{28}, \bibinfo{number}{1} (\bibinfo{year}{1972}), \bibinfo{pages}{11--21}.
\newblock
\urldef\tempurl%
\url{https://doi.org/10.1108/eb026526}
\showDOI{\tempurl}


\bibitem[Tang and Yang(2024)]%
        {Tang2024MultiHopRAGBR}
\bibfield{author}{\bibinfo{person}{Yixuan Tang} {and} \bibinfo{person}{Yi Yang}.} \bibinfo{year}{2024}\natexlab{}.
\newblock \showarticletitle{MultiHop-RAG: Benchmarking Retrieval-Augmented Generation for Multi-Hop Queries}.
\newblock \bibinfo{journal}{\emph{CoRR}}  \bibinfo{volume}{abs/2401.15391} (\bibinfo{year}{2024}).
\newblock
\urldef\tempurl%
\url{https://doi.org/10.48550/ARXIV.2401.15391}
\showDOI{\tempurl}
\showeprint[arXiv]{2401.15391}


\bibitem[Traag et~al\mbox{.}(2019)]%
        {Traag_2019}
\bibfield{author}{\bibinfo{person}{V.~A. Traag}, \bibinfo{person}{L. Waltman}, {and} \bibinfo{person}{N.~J. van Eck}.} \bibinfo{year}{2019}\natexlab{}.
\newblock \showarticletitle{From Louvain to Leiden: guaranteeing well-connected communities}.
\newblock \bibinfo{journal}{\emph{Scientific Reports}} \bibinfo{volume}{9}, \bibinfo{number}{1} (\bibinfo{date}{March} \bibinfo{year}{2019}).
\newblock
\showISSN{2045-2322}
\urldef\tempurl%
\url{https://doi.org/10.1038/s41598-019-41695-z}
\showDOI{\tempurl}


\bibitem[Yang et~al\mbox{.}(2018)]%
        {yang2018hotpotqa}
\bibfield{author}{\bibinfo{person}{Zhilin Yang}, \bibinfo{person}{Peng Qi}, \bibinfo{person}{Saizheng Zhang}, \bibinfo{person}{Yoshua Bengio}, \bibinfo{person}{William~W. Cohen}, \bibinfo{person}{Ruslan Salakhutdinov}, {and} \bibinfo{person}{Christopher~D. Manning}.} \bibinfo{year}{2018}\natexlab{}.
\newblock \showarticletitle{{HotpotQA}: A Dataset for Diverse, Explainable Multi-hop Question Answering}. In \bibinfo{booktitle}{\emph{Conference on Empirical Methods in Natural Language Processing ({EMNLP})}}.
\newblock


\end{thebibliography}
}

\appendix

\section{Validation of Generated Statements}

We evaluated statement extraction faithfulness by sampling 1,000 statements per dataset and comparing each one to the original text block. Table~\ref{tab:statement_alignment} reports the proportion of accurate statements for general and tabular domains. A statement is considered accurate if it preserves the intended meaning of the original chunk without introducing errors such as hallucinations.

\begin{table}[ht]
\centering
\caption{Alignment of generated statements against the original chunked text.}
\label{tab:statement_alignment}
\renewcommand{\arraystretch}{1}
\setlength{\tabcolsep}{8pt}      
\begin{tabular}{@{}llc@{}}
\toprule
\textbf{Domain}           & \textbf{Dataset}      & \textbf{Statement Accuracy} \\
\midrule
\multirow{3}{*}{General}  & MultiHop-RAG          & 96.5\% \\
                          & ConcurrentQA          & 97.4\% \\
                          & WikiHowQA             & 97.9\% \\
\cmidrule(lr){2-3}
                          & \textit{Average}      & \textit{97.3\%} \\
\midrule
\multirow{3}{*}{Tabular}  & SEC-10Q               & 94.3\% \\
                          & NTSB                  & 97.7\% \\
\cmidrule(lr){2-3}
                          & \textit{Average}      & \textit{96.0\%} \\
\bottomrule
\end{tabular}
\end{table}

Statement generation attains \(\geq 96\%\) accuracy, slipping only slightly on table-centric sources. This confirms that our LLM-proposition extractor still delivers high precision. SEC-10Q tables pose extra hurdles, such as scattered units/periods and repetitive layouts that split related details and blur entity links. The tabular enhancement step (Section \ref{subsec:statement_enhancement_tabular}) restores context by attaching headers, units, and other metadata to each extracted proposition.

\section{HLG Structure Analysis}
Table~\ref{tab:dataset_node_counts} provides an overview of the node distribution within HLG across multiple datasets.

\begin{table}[ht]
  \setlength{\abovecaptionskip}{4pt}     
  \setlength{\belowcaptionskip}{-2pt}
\centering
\caption{Number of nodes in each tier of HLG for each dataset.}
\label{tab:dataset_node_counts}
\renewcommand{\arraystretch}{1}
\setlength{\tabcolsep}{2pt}     
\begin{tabular}{@{}lcccccc@{}}
\toprule
\textbf{Dataset} & \textbf{Source} & \textbf{Chunk} & \textbf{Topic} & \textbf{Statement} & \textbf{Fact} & \textbf{Entity} \\
\midrule
MultiHop-RAG     & 609     & 6,272   & 7,067   & 53,927   & 274,890   & 18,291 \\
SEC-10Q          & 20      & 5,054   & 5,219    & 47,284   & 57,443    & 6,903  \\
NTSB             & 20      & 6,574   & 7,633    & 86,837   & 50,462    & 8,867  \\
WikiHowQA        & 5000    & 15,255  & 15,455   & 163,456  & 174,262   & 36,997 \\
ConcurrentQA     & 13,500  & 55,328  & 57,213   & 495,835  & 558,897   & 106,859 \\
\bottomrule
\end{tabular}
\end{table}

\subsection{Scalability and Performance}
\label{subsec:scalability_performance}

\subsubsection{Indexing Performance (MultiHop-RAG dataset):}
The dataset can be indexed and ingested in under 1 hour using AWS Neptune Analytics for graph storage and AWS OpenSearch for vector storage. Token consumption for LLM-based extraction (\texttt{Claude-3 Sonnet}):

\begin{table}[h!]
  \setlength{\abovecaptionskip}{4pt}   
  \setlength{\belowcaptionskip}{-2pt}    
  \centering
  \renewcommand{\arraystretch}{0.8}      
  \caption{Token statistics during indexing}
  \begin{tabular}{l@{\hspace{0.8em}}r@{\hspace{0.8em}}r}
    \toprule
    \textbf{Stage} & \textbf{Input Tokens} & \textbf{Output Tokens}\\
    \midrule
    Statement Extraction          & 3{,}098{,}745 & 1{,}827{,}175\\
    Topics/Entities/Relationships & 10{,}812{,}763 & 5{,}080{,}192\\
    \midrule
    \textbf{Total}                & 13{,}911{,}508 & 6{,}907{,}367\\
    \bottomrule
  \end{tabular}
  \label{tab:indexing-tokens}
\end{table}

Under AWS Bedrock prices (December~2024), end-to-end indexing with \texttt{Claude-3~Sonnet} for LLM inference and \texttt{Cohere Embed-English-v3} for embeddings costs \$145.34.  
Running the same workload in batch mode halves the LLM expenditure to \$72.67 without affecting quality, and the embedding stage adds only \$0.22. Since boilerplate text (e.g., legal disclaimers) recurs across articles, enabling caching lowers token counts and total spend even further.

\subsubsection{Retrieval Performance} On a standard AWS \texttt{g5.xlarge} instance, an uncached query, covering passage retrieval, embedding creation, and LLM call(s), completes in 20–30 s and costs \$0.032.  
Once the query’s vectors and LLM outputs are cached, the same request is answered in well under one second with no additional charge, meeting real-time requirements while keeping operating costs minimal. Future work includes methods to recognize single-hop queries to avoid unnecessary multi-hop expansions, further reducing costs and latency (Section~\ref{sec:discussion_future}).

\section{Limitations and Future Work}
\label{sec:discussion_future}

\subsection{Limitations}
\subsubsection{LLM Invocation and Indexing Costs}

HLG delivers high-fidelity extractions, but at the cost of multiple LLM calls. Each chunk is processed twice; first to extract statements, then to refine topics and entities, so corpora with hundreds of millions of tokens accumulate substantial compute. These costs arise from building a fine-grained knowledge graph that explicitly encodes topics, entities, and relations. Expenses can be reduced by delegating the statement extraction step to a lighter model such as \emph{wiki-flan-t5-large} (783 M parameters) with little loss in retrieval quality. Because extraction is performed once offline during index construction, online queries add no further LLM latency.

\subsubsection{Single-Hop Domains and Early Stopping}
\label{subsubsec:single-hop}
Our evaluation of WikiHowQA highlights a minor regression for multi-hop graph methods in single-hop contexts. 
When the query addresses a single chunk of text, graph expansions can introduce extraneous statements or prolong retrieval. 
An early-stopping heuristic could detect queries dominated by a single source document and return top-$k$ statements directly from that source, improving efficiency and reducing noise.

\subsubsection{LLM Beam Search vs.\ Lightweight Reranking}
An early alternative to our BGE-based reranker was an LLM-driven beam search, where each expansion step invoked a language model to rank expansions. This approach showed strong semantic alignment but was prohibitively slow for practical use.  Swapping to a lighter reranker maintained strong performance gains while lowering latency, making graph-based methods viable in production environments.

\subsection{Error Analysis}
We observe several recurring error modes:

\subsubsection{Over-Expansion} 
Even with beam size limited to 3 expansions, queries with highly connected entities (i.e., supernodes) can cause the pipeline to retrieve loosely related statements. Our entity-overlap ranking mitigates this by surfacing neighbours that share more than one entity, but some out-of-scope expansions remain.

\subsubsection{Numeric/Tables Misalignment} 
In SEC-10Q, numeric values sometimes appear in multiple contexts. If two statements are semantically similar but reference different quarters, the top-$k$ filtering can inadvertently retrieve the wrong time period. This is especially important for statement-level tabular enhancements (Section \ref{subsec:statement_enhancement_tabular}).

\subsubsection{Duplicate or Near-Duplicate Statements}
The same or nearly identical facts can appear across different documents or versions (especially in historical filings). Without diversity filtering, these duplicates can dominate the top-$k$ and reduce coverage. Our 0.5\% threshold helps alleviate this but does not fully eliminate overlapping data for heavily repeated claims.

\subsection{Future Work}

\subsubsection{Propositionizer Pipelines}
Subsequent research will prioritize the development of more efficient propositionization, reducing indexing tokens and LLM calls. 
Aligning the propositionizer’s output with our knowledge-graph schema (topics/entities/relationships) remains key to maintaining high-quality links.

\subsubsection{Hybrid Retrieval of Chunks and Statements}
We plan to explore a layered pipeline that first employs chunk-level searches to localize relevant segments, followed by statement-level expansions or reranking for multi-hop clarity.  Such a hybrid approach might preserve the speed of chunk-based retrieval while leveraging more precise graph-based statements where necessary.

\subsubsection{Domain Specialization and Fine-Tuning}
Enhanced domain adaptation, particularly within finance, healthcare, and legal corpora, may sharpen the extraction of domain-specific topics, entities, and relations. Targeted fine-tuning of the reranker and propositionizer components in these specialized settings is expected to improve precision while keeping inference costs manageable. Although we deliberately avoided highly specialized models in this work to ensure broad applicability with general-purpose LLMs, we are internally testing fine-tuned approaches for structured numeric data (e.g., unit conversion tools, temporal linking improvements). These are ongoing efforts to refine proposition extraction in complex, table-based contexts like legal or financial search.

\end{document}